\documentclass{elsart1p}

 \usepackage{graphicx}

\usepackage{amssymb}

\def \bracket<#1>{\mbox{$\langle {#1}\rangle$}}
\def \mvec #1{\mbox{\boldmath{${#1}$}}}
\def \mvecsc #1{\mbox{\scriptsize \boldmath{${#1}$}}}
\def \ketv #1>{\mbox{$|{#1}\rangle$}} 
\def \mate<#1|#2|#3>{\mbox{$\langle {#1}|\,{#2}\,|{#3}\rangle$}}
\def \etal{{\it et\,al.\,\,}}
\def \ie{{\it ie.\,\,}}

\def\YTD(#1,#2){%
\unitlength=6pt%
\begin{picture}(#1,2)(0,0)%
\multiput(0,0)(1,0){#2}{\framebox(1,1)}%
\multiput(0,1)(1,0){#1}{\framebox(1,1)}%
\end{picture}}

\def\YTT(#1,#2,#3){%
\unitlength=6pt%
\begin{picture}(#1,3)(0,1)%
\multiput(0,0)(1,0){#3}{\framebox(1,1)}%
\multiput(0,1)(1,0){#2}{\framebox(1,1)}%
\multiput(0,2)(1,0){#1}{\framebox(1,1)}%
\end{picture}}

\begin{document}

\begin{frontmatter}



\title{A pentaquark picture of $\Lambda(1405)$}


\author{Takashi Inoue}

\address{Dept. Phys. Sophia University\\ 7-1 Kioi-cho, Chiyoda-ku, Tokyo 102-8554, Japan}

\begin{abstract}
We test a pentaquark model of the negative parity hyperon $\Lambda(1405)$.
We use a model where valence four quarks and one antiquark move in a potential relativistically 
and interact with each other and themselves through gluon and NG boson.
Fall-apart decay rate of the system is studied and strong model dependence is pointed out.
\end{abstract}

\end{frontmatter}

\section{Introduction}
\label{}
The negative parity hyperon $\Lambda(1405)S_{01}$
is interested for many years and several interpretations are proposed.
The most simplest one is so called P-wave excited 3-quark system.
Such system is studied, for example, 
by Isgur and Karl using harmonic-oscillator and one-gluon-exchange perturbation,
and by Furuichi \etal as a bound state problem. 
It turns out that to reproduce mass of $\Lambda(1405)$ is difficult in this approach.
A lattice study shows that the flavor singlet 3-quark system
with $I(J^P) = 0(1/2^-)$, cannot be so light as $\Lambda(1405)$. 
The most popular scenario today is bound $\bar K N$ system 
or dynamically generated resonance in meson-baryon scattering. 
In this decade, mason-baryon coupled channel scattering are studied extensively.
There, several baryons including $\Lambda(1405)$, are generated dynamically
as a resonance or a quasi bound state which decay to open channels.
Most of studies based on the chiral Lagrangian and the Bethe-Salpeter equation.

Besides these scenario, pentaquark is a possible interpretation of the hyperon.
Here, pentaquark stands for a state with five valence constituents.
In other words, it's quark number decomposition starts form $\ketv qqqq \bar q>$ states.
By today, we do not have established pentaquark.
Indeed $\Theta^+$ is a good candidate though it's existence is still questionable.
Therefore, if $\Lambda(1405)$ is well described as a pentaquark, 
we encounter exotic pentaquark baryon for the first time.

\section{Pentaquark model}%
To make negative parity baryon with four quarks and one anti-quark,
it is simplest to put all quarks and antiquark into the ground S-wave orbit, 
namely S-shell pentaquark. 
The whole system should be color singlet for the QCD quark confinement.
On top of that, 4-quark subsystem must be totally antisymmetric from the Fermi statistics. 
There are several flavor-spin configuration satisfying these conditions.
We consider three configurations where 4-quark subsystem is combined to
$(\, {\bf [2,1,1]_f} ~,~ \bf{ [3,1]_s } \,)$ and
$(\, {\bf [2,1,1]_f} ~,~ \bf{ [2,2]_s } \,)$ and
$(\, {\bf [2,2]_f  } ~,~ \bf{ [3,1]_s } \,)$ in flavor and spin space respectively.
With the $\bf[211]_f$ subsystem, pentaquark is flavor singlet with an antiquark.
While, the $\bf [22]_f$ one leads flavor octet and anti-decuplet pentaquarks,
where one $\Lambda$ hyperon exist at center of octet.
In all, we have three $\Lambda(1/2^-)$ pentaquarks, two flavor single and one octet.
In the conventional 3-quark approach, $\Lambda(1405)$ is flavor singlet.
While, in the chiral unitary model, one pole corresponding to $\Lambda(1405)$
is almost flavor octet\cite{Jido:2002yz}.

We define unperturbed pentaquark states as
$\ketv B >^0 = u_0(x_1) u_0(x_2) u_0(x_3) u_0(x_4) \bar v_0(x_5)$
a product of the ground states of valence quark localized by a potential $S(r) - \gamma^0 V(r)$,
and introduce a residual interaction 
${\cal L}_I = - \bar{\psi}(x) \left\{ 
     g_s \gamma^\mu A_\mu^a(x) \frac{\lambda^a}{2} +
     S(r) i \gamma^5 \frac{\hat \Phi (x)}{F} 
                \right\}\psi(x)$
by which valence (anti-)quark emits and absorbs
perturbative gluon and octet pseudoscalar meson or NG-bosons.
Here, the quark-meson coupling term is determined so that it recovers the chiral symmetry
broken by the scalar potential $S(r)$. 
We set up perturbation 
$ \mate<B|{\cal O}|B> = 
 {}^0\!\langle B| \sum \frac{i^n}{n!}\! 
 \int \! \! d^3 \vec x \, d^4 x_1 \cdots d^4 x_n 
 T[ { {\cal L}_{I}(x_1) \cdots {\cal L}_{I}(x_n) }\, 
    {\cal O}(\vec x) \, ]
 | B \rangle^0 $
where baryon matrix element of a local operator is calculated
with the unperturbed state and the interaction.
Thus, quark correlation and supplementing meson cloud \ie extra quarks in baryon, are introduced.
This formalism is developed at T\"ubingen originally for nucleon,
and we call it the perturbative chiral quark model\cite{Lyubovitskij:2000sf}.

We've tested this pentaquark picture of $\Lambda(1405)$ 
by studying magnetic moment, mass shift and $\sigma$-term.
It turns out there is two pentaquark $\Lambda(1405)$ almost degenerated,
one flavor singlet and one flavor octet.
This feature is common with the chiral unitary model.
In this paper, we study decay rate of the system.

\section{Decay of S-shell Pentaquark}
Pentaquark can decay via a characteristic mechanism called the fall-apart
where it simply separate into baryon$(qqq)$ and meson$(q\bar q)$,
namely baryon emit meson without creating $q\bar q$.
Hosaka \etal studied $\Theta^+ \to (K N)_{I=0}$ 
decay via this mechanism\cite{Hosaka:2004bn}.
They obtained $\Gamma_{\Theta^+} \simeq {890}$ MeV for $1/2^-$ configuration
and conclude that $\Theta^+$ cannot be $1/2^-$.
We follow the paper and calculate the decay rate of $\Lambda(1405) \to (\Sigma \pi)_{I=0}$. 

The decay rate is given by 
$\Gamma_{\Lambda} 
  = \frac{M_{\Sigma}(E_{\Sigma} + M_{\Sigma})}{E_{\Sigma} M_{\Lambda}} 
    \frac{q}{2\pi} \left| {\cal M}_{\Lambda \to \Sigma^- \pi^+} \right|^2$
where $q$ is released momentum with $q \simeq 150.4$ MeV.
To reproduce empirical decay rate of $\Gamma_{\Lambda} \simeq 50$ MeV, 
the transition amplitude ${\cal M}$ should be around $1.2$.
This amplitude is determined by the interaction Lagrangian with
${\cal M} = \mate<\Sigma^- \pi^+|\int \! d^3 \mvec x \, {\cal L}_{I}(\mvec x)|\Lambda>$.
It is calculated by 
\begin{equation}
{\cal M} \simeq 2 \sqrt2 \, a \! \int\!\! d^3 \mvec x \, 
    {S(|\mvec x|)}/{F} \, \mate<0| \bar\psi(\mvec x) \gamma_5 \psi(\mvec x)|(u\bar d)^{\pi^+}>\,
    e^{-i \mvecsc q \cdot \mvecsc x}
\label{eqn:amp}
\end{equation}
where $a$ is the spectroscopic factor \ie the amplitude with which 
pentaquark $\Lambda(1405)$ contain $\Sigma^- \pi^+$ component,
namely $\ketv \Lambda(1405)_{1_f/8_f}>= a \ketv (dds)^{\Sigma^-} (u\bar d)^{\pi^+}> + \cdots$.
We obtain $a = {\pm 1}/{(4\sqrt2)}$ for $1_f/8_f$ respectively 
by using explicit pentaquark wave function.

Following the paper by Hosaka \etal, we borrow a non-relativistic Gaussian wave function
$\phi(\mvec r_i) = \left({\alpha_0^2}/{\pi}\right) 
                   e^{-\frac{\alpha_0^2}{2}{\mvecsc r_i}^2}$
with $\alpha_0^2 = 3.0$ {fm}$^{-2}$ for valence quark.
With this, the relative and c.m. coordinates of $q \bar q$ 
are easily separated and we obtain 
\begin{equation}
\mate<0|\bar\psi(\mvec x) \gamma_5 \psi(\mvec x)|(u\bar d)^{\pi^+}>
   \simeq \sqrt2 \left({\alpha_0}/{\pi}\right)^{\frac34} \,
          \left({\alpha}/{\pi}\right)^{\frac34}  \, e^{-\frac{\alpha^2}{2}\mvecsc x^2}
\label{eqn:mat}
\end{equation}
with $\alpha^2 = 6/5 \, \alpha_0^2$. 
We calculate the amplitude eq.(\ref{eqn:amp}) using the matrix element eq.(\ref{eqn:mat}).

\section{Results and Discussion}%

\begin{figure}[t]
 \centering
 \includegraphics[width=0.40\textwidth]{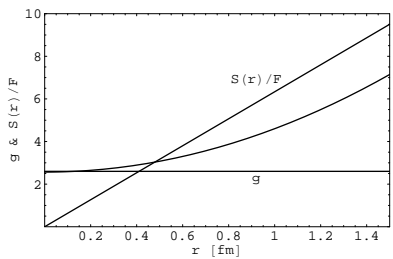}
 \caption{The several quark-meson couplings used in this study.}
 \label{fig:couplings}
\end{figure}

First, we use the typical linear scalar potential 
$S(r) = c r$ with $c=0.11$ GeV$^2$ and low energy constant $F=88$ MeV.
The coupling $S(r)/F$ is shown in Fig. \ref{fig:couplings}.
The amplitude is obtained as ${\cal M} \simeq 4.9$,
which is four times larger than one of empirical data,
and gives much large decay rate of $\Gamma \simeq 950$ MeV.
We need to reconsider choice of potential to localize valence quark.
In the original perturbative chiral quark model for nucleon,
a quadratic potential $S(r)$ and $V(r)$ are used. 
That potentials are essentially same as the linear $S(r)$, for nucleon, 
although they don't give negative energy bound states.
The coupling in this case is show in Fig. \ref{fig:couplings}.
With this coupling, the amplitude is still large value of 4.0 yet.
In the paper by Hosaka \etal, a conventional constant coupling $g = 2.6$ is used. 
If we use this coupling, the amplitude becomes $2.5$,
which is one half of the first result and gives $\Gamma \simeq 250$ MeV. 
The result is larger than data but not so far from it. 

We see that the size of fall apart decay rate depends strongly 
on the choice of the quark-meson coupling.
From the chiral symmetry point of view,
the coupling should be given by the scalar potential to localize valence quarks.
However, such couplings give large decay rate as shown above.
Weaker $S(r)$ with supplementing $V(r)$ can give reasonable amplitude,
however we know that $S(r)$ is important to produce meson cloud
for nucleon property, for example $\pi$-N $\sigma$-term.
The matrix element in eq(\ref{eqn:mat}) ignore small component of quark wave function.
Taking small component may change the results considerably. 



\end{document}